\documentclass[conference]{IEEEtran}
\IEEEoverridecommandlockouts
\usepackage{amsmath,amsfonts, nicematrix}
\usepackage{cuted}
\usepackage{cite}
\usepackage{bm}
\usepackage{amsmath,amssymb,amsfonts}
\usepackage{algorithmic}
\allowdisplaybreaks[4]
\usepackage{graphicx}
\usepackage{textcomp}
\usepackage{xcolor}
\usepackage{booktabs}
\usepackage{makecell}
\usepackage{subfigure}
\usepackage[colorlinks,linkcolor=blue,citecolor=blue,urlcolor=black]{hyperref}
\usepackage{fancyhdr}

\newcounter{MYtempeqncnt}
\def\BibTeX{{\rm B\kern-.05em{\sc i\kern-.025em b}\kern-.08em
    T\kern-.1667em\lower.7ex\hbox{E}\kern-.125emX}}
\begin{document}

\title{Resilience-oriented Planning and Cost Allocation of Energy Storage Integrated
with Soft Open Point Based on Resilience Insurance
\thanks{This work was supported by the National Natural Science Foundation of China (U22B20103, 52307137). Corresponding author: Gengfeng Li.}
}

\author{Bingkai Huang, Yuxiong Huang, Qianwen Hu, Gengfeng Li, Zhaohong Bie  \\
\textit{School of Electrical Engineering}\\ \textit{Xi’an Jiaotong University} \\ Xi'an, China  \\
bingkaihuang@stu.xjtu.edu.cn, yuxionghuang@xjtu.edu.cn, qwhu089@stu.xjtu.edu.cn, \\ gengfengli@xjtu.edu.cn, zhbie@mail.xjtu.edu.cn
}

\maketitle

\begin{abstract}
In recent years, frequent extreme events have put forward higher requirements for improving the resilience of distribution networks (DNs). Introducing energy storage integrated with soft open point (E-SOP) is one of the effective ways to improve resilience. 
However, the widespread application of E-SOP is limited by its high investment cost. 
Based on this, we propose a cost allocation framework and optimal planning method of
E-SOP in resilient DN. Firstly, a cost allocation mechanism for E-SOP based on resilience insurance service is designed; the probability of power users purchasing resilience insurance service is determined based on the expected utility theory. 
Then, a four-layer stochastic distributionally robust optimization (SDRO) model is developed for E-SOP planning and insurance pricing strategy. In this model, the uncertainty in the intensity of contingent extreme events is addressed using a stochastic optimization approach. Meanwhile, the uncertainties in outage occurrences and resilience insurance purchases, which result from a specific extreme event, are handled through a distributionally robust optimization approach.
Finally, the effectiveness of the proposed model is verified on the modified IEEE 33-bus DN.
\end{abstract}


\thispagestyle{fancy}
\fancyhf{}
\renewcommand{\headrulewidth}{0pt}
\fancyhead[C]{\footnotesize
\textcopyright 2025 IEEE. Personal use permitted. For other uses, permission required. 
This paper has been accepted and published in 2025 IEEE PESGM.
 
This is author's accepted manuscript, which may differ from final version. 
Final version available at IEEE Xplore: https://ieeexplore.ieee.org/document/11225605, DOI: 10.1109/PESGM52009.2025.11225605}

\section{Introduction}
In recent years, frequent extreme events, such as natural disasters and man-made attacks, have brought significant challenges to distribution network (DN) operations and seriously harmed energy security and social development. The traditional DN, which mainly takes reliability as the planning standard, is challenging to deal with the impact of extreme events. Consequently, it is imperative to improve the resilience of DN \cite{bielin}.

Energy storage integrated with soft open points (E-SOPs) offers a promising solution for enhancing the resilience of DNs \cite{yangxuwen}. 
As a power electronic device in place of the normally open point (NOP), E-SOP swiftly restores the loads at the end of the feeder by transferring power from the energy storage system (ESS) and other feeders and providing reactive power support when the faults occur \cite{liji}. 
Compared with transitional normally open point (NOP), E-SOP can be powered by ESS, overcoming the defect that traditional NOP requires at least one feeder not to fail during power transfer \cite{yangxu}. Besides, compared with transitional ESS connected to the system through a single bus, E-SOP also allows ESS to be accessed across multiple buses in the network, facilitating the sharing power of ESS among multiple feeders \cite{lvliang}. In a word, E-SOP further improves the resilience of DN. 

However, E-SOPs are quite costly at present, necessitating optimal planning and cost allocation methods to promote the application of E-SOPs in resilient DN. There has been considerable research on the planning methods for ESSs, SOPs and E-SOPs in DNs. A cooperative planning model of mobile ESS and microgrids was established in \cite{kimdvorkin}. Ref. \cite{yangzhou} proposed a co-deployment framework for SOPs and remote-controlled switches to improve resilience in DN. A sequential optimization model of active DN with E-SOP was proposed in \cite{yaozhou}. There have been some studies on the cost allocation mechanism and insurance in DN. Ref. \cite{liuwu} proposed the concept of dispatch insurance, which leverages the financial interests of the utility and market participating DERs to resolve the conflict issue. Ref. \cite{xuyang} designed deviation insurance for new energy sources to realize the cost allocation of shared ESS.

Based on the above research, this paper studies E-SOP's cost allocation mechanism and optimal planning method in resilient DN. The main contributions of this paper are as follows:
\begin{itemize}
\item We design a cost allocation framework for E-SOP based on resilience insurance, and establish the probability model of power users purchasing resilience insurance services under different levels of extreme events.

\item We establish a four-layer optimal model using stochastic distributionally robust optimization (SDRO) to determine the optimal planning scheme of E-SOP and insurance pricing strategy.
\end{itemize}

In the remainder of this paper, Section \ref{sec: Cost Allocation Framework Based on Resilience Insurance} describes the cost allocation framework based on resilience insurance; Section \ref{sec: SDRO Model for E-SOP Planning} establishes the E-SOP optimal planning model using SDRO method; case study and conclusions are given in Section \ref{sec: Case Study} and Section \ref{sec: Conclusion}.

\section{Cost Allocation Framework Based on Resilience Insurance}
\label{sec: Cost Allocation Framework Based on Resilience Insurance}
\subsection{E-SOP Resilience Insurance Design}
The users in the DN can be classified into three levels based on their importance. Level-1 users include critical loads where power interruptions could directly result in personal injury, loss of life, or significant economic damage, such as government agencies and airport facilities. Level-2 users are less critical than level-1 but still experience substantial economic losses from power outages, encompassing large industrial facilities and other key infrastructure. Level-3 users are non-critical consumers, such as residential buildings and non-essential industries, whose power supply can be interrupted without severe consequences during extreme events and who are not potential customers for the resilience insurance.

\begin{figure}[ht]
\centerline{\includegraphics[width=0.9\columnwidth]{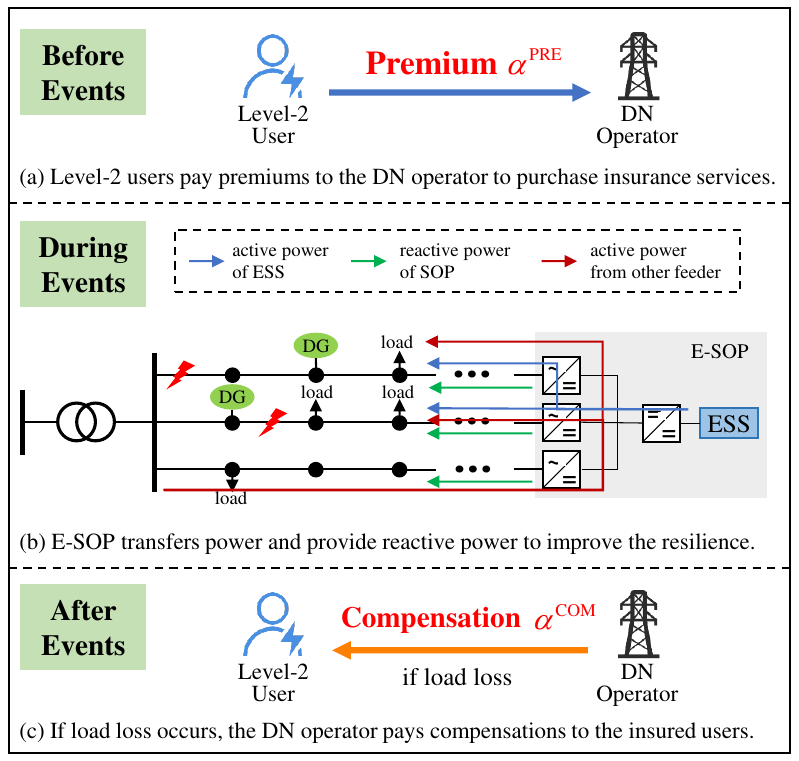}}
\caption{Schematic diagram of the proposed resilience insurance.}
\label{fig: E-SOP improve resilience of the distribution network}
\end{figure}

It is often considered to be the obligation of the DN operator to improve supply resilience for level-1 users under extreme events. DN operators will only consider the supply of level-2 users on the premise of ensuring the supply of level-1 users. Some level-2 users may also need to increase resilience, and this part of the level-2 users is the target customers of E-SOP resilience insurance services. This paper proposes a resilience insurance based on E-SOP, as shown in Fig.\ref{fig: E-SOP improve resilience of the distribution network}.

As shown in Fig.\ref{fig: E-SOP improve resilience of the distribution network}, the level-2 user purchases resilience insurance from the DN operator, and the DN operator receives the premium. According to the principle of “who benefits, who bears,” the DN operator should ensure the power supply of insured level-2 users as much as possible during the extreme events by transferring power from ESS and other feeders and providing reactive power using E-SOP. In other words, the insured level-2 user is regarded as level-1 users during the schedule. On the contrary, if the DN operator fails to ensure the power supply of the insured level-2 user, the DN operator will pay compensation according to the contract.
DN operators can cover the investment cost of E-SOP by collecting premiums, while insurance customers can save the cost of configuring a backup power supply by purchasing insurance and ensuring the power supply. In summary, the proposed resilience insurance service realizes the cost allocation of E-SOP.

\subsection{Probability of Users Purchasing Resilience Insurance}
The probability of users purchasing resilience insurance services before extreme events $\rho _{s,k}^{{\rm{load}}}$ can be obtained by expected utility theory, and the specific model is as follows.

Before making decisions, level-2 users have the following prior information: the loss of load probability (LOLP) during extreme event $s$ is $\overline p^{{\rm{1st}}}_s$ for level-1 users, and $\overline p^{{\rm{2nd}}}_s$ for level-2 users; while the expected energy not served (EENS) during extreme event $s$ is ${\overline E^{{\rm{1st}}}_s}$ for level-1 users, and ${\overline E^{{\rm{2nd}}}_s}$ for level-2 users. Additionally, following assumptions are proposed for mathematical simplicity and tractability:

\noindent \textbf{Assumption 1.} If a level-2 user has insurance, he will have the same insurance priority in the scheduling process as a level-1 user. Therefore, when level-2 users decide whether to purchase insurance, they will use the relevant prior data of level-1 users to calculate their expected utility after purchasing insurance.

\noindent \textbf{Assumption 2.} The energy not served (ENS) during extreme events follows a normal distribution with a mean of EENS and a variance set to a value $\Sigma^2$, i.e.,:
\begin{equation}    \label{eq: normal distribution}
    E_{s,k}^{\rm{1st}} \sim N \left( \overline E^{{\rm{1st}}}_s, \Sigma^2_1 \right), \ E_{s,k}^{\rm{2nd}} \sim N \left( \overline E^{{\rm{2nd}}}_s, \Sigma^2_1 \right)
\end{equation}
where the subscript 1st and 2nd represent the user's purchase and non-purchase of insurance, respectively; $E_{s,k}^{\bullet}$ denotes ENS under scenario $s$.

\noindent \textbf{Assumption 3.} We adopt quadratic loss functions to characterize the users' losses due to energy not served:
\begin{align}
    & u_{s,k}^{\rm{1st}} = {l_{k}} (E_{s,k}^{\rm{1st}})^2 - \alpha^{\rm{COM}}_s E_{s,k}^{\rm{1st}} \\
    & u_{s,k}^{\rm{2nd}} = {l_{k}} (E_{s,k}^{\rm{2nd}})^2
\end{align}
where $u_{s,k}^{\bullet}$ represents users loss; $l_{k}$ represents quadratic loss function coefficient; $\alpha^{\rm{COM}}_s$ is the compensation per unit of electricity. 

Obviously, user $k$'s action set $\mathcal{A}_{s,k}$ under scenario $s$ consists of two elements: purchasing insurance $a_{s,k}^{{\rm{1st}}}$ and not purchasing $a_{s,k}^{{\rm{2nd}}}$, i.e., $\mathcal{A}_{s,k} = \{ a_{s,k}^{{\rm{1st}}}, a_{s,k}^{{\rm{2nd}}} \}$. The state set $\Theta_{s,k}$ also contains two elements: energy supplied $B_{s,k,t}^{{\rm{load,loss}}} = 0$ and energy not supplied $B_{s,k,t}^{{\rm{load,loss}}} = 1$. Therefore, the gain matrix of user $k$ in extreme event $s$ is as follow:
\begin{equation}
    G_{s,k} = \begin{pNiceMatrix}[first-row,last-col]
	a_{s,k}^{{\rm{1st}}} & a_{s,k}^{{\rm{2nd}}} &   \\
	- C_{s,k} - u_{s,k}^{\rm{1st}} & - u_{s,k}^{\rm{2nd}} & B_{s,k,t}^{{\rm{load,loss}}} = 1 \\
	- C_{s,k} & 0 & B_{s,k,t}^{{\rm{load,loss}}} = 0 \\
    \end{pNiceMatrix}
\end{equation}
where $C_{s,k}$ is the total premium. In this paper, we assume that the insurance premium per kWh $\alpha^{\rm{PRE}}_s$ is determined by the system operator. The premium $C_{s,k}$ is the product of $\alpha^{\rm{PRE}}_s$ and the user's estimated electricity consumption, i.e., $C_{s,k} = \alpha^{\rm{PRE}}_s \sum\nolimits_{t=1}^T P_{s,k,t}^{\rm{load}} \Delta T$.

Note that the price of insurance services should be subject to the rule that the loss of the user who has purchased insurance is always less than the loss of the underinsured if the energy is not supplied during the extreme events, i.e., $ C_{s,k} + u_{s,k}^{\rm{1st}} \le u_{s,k}^{\rm{2nd}}$. Thus, the loss matrix is as follow:
\begin{equation}
    L_{s,k} = \begin{pNiceMatrix}[first-row,last-col]
	a_{s,k}^{{\rm{1st}}} & a_{s,k}^{{\rm{2nd}}} &   \\
	0 & u_{s,k}^{\rm{2nd}} - u_{s,k}^{\rm{1st}} - C_{s,k} & B_{s,k,t}^{{\rm{load,loss}}} = 1 \\
	C_{s,k} & 0 & B_{s,k,t}^{{\rm{load,loss}}} = 0 \\
    \end{pNiceMatrix}
\end{equation}

Therefore, the average losses $E_{\rm{loss}}(a_{s,k}^{{\rm{1st}}})$ and $E_{\rm{loss}}(a_{s,k}^{{\rm{2nd}}})$ for user $k$ making different decisions are as follows:
\begin{align}
    & E_{\rm{loss}}(a_{s,k}^{{\rm{1st}}}) = C_{s,k} (1 - \overline p^{{\rm{1st}}}_s) \\
    & E_{\rm{loss}}(a_{s,k}^{{\rm{2nd}}}) = (u_{s,k}^{\rm{2nd}} - u_{s,k}^{\rm{1st}} - C_{s,k}) \overline p^{{\rm{2nd}}}_s
\end{align}

If $E_{\rm{loss}}(a_{s,k}^{{\rm{1st}}}) \le E_{\rm{loss}}(a_{s,k}^{{\rm{2nd}}})$, the user $k$ will buy insurance. Thus, the probability of user $k$ buying insurance is as follows:
\begin{equation}    \label{eq: rho load}
    \rho_{s,k}^{{\rm{load}}} = \Pr \{ E_{\rm{loss}}(a_{s,k}^{{\rm{1st}}}) \le E_{\rm{loss}}(a_{s,k}^{{\rm{2nd}}})\}
\end{equation}
When the premium $\alpha^{\rm{PRE}}_s$ and compensation $\alpha^{\rm{COM}}_s$ of the insurance contract are given, the probability $\rho_{s,k}^{\rm{load}}$ of user $k$ buying insurance under scenario $s$ is a definite value, which can be determined by solving \eqref{eq: rho load}. In other words, $\rho_{s,k}^{\rm{load}}$ is a function of $\alpha^{\rm{PRE}}_s$ and $\alpha^{\rm{COM}}_s$, i.e., $\rho_{s,k}^{\rm{load}} = \rho_{s,k}^{\rm{load}}(\alpha^{\rm{PRE}}_s, \alpha^{\rm{COM}}_s)$. Please note that decision makers can replace Eq. \eqref{eq: normal distribution}-\eqref{eq: rho load} with a more complex user behavior model.

\section{Four-Layer SDRO Model for E-SOP Planning and Insurance Pricing}
\label{sec: SDRO Model for E-SOP Planning}
\subsection{Ambiguity Set}
\label{subsec: Ambiguity Set}
The willingness of users to purchase insurance before an extreme event occurs and the outage of lines during an extreme event are key factors affecting the planning and operation of E-SOP. Thus, the ambiguity set of coupling between line outages fault and user insurance intention is defined as follows:
\begin{equation} \label{eq: Ambiguity Set}
    \mathcal{B} = \left\{  
    \begin{array}{ll}
        \mathbb{P}_s \in \mathcal{P}(\Xi_s) \left\lvert\, \begin{array}{l}
            \mu _{s,mn,t}^{{\rm{line}}}, \ \mu _{s,k}^{{\rm{load}}} \in \Xi_s \\
            \text{E}_{\mathbb{P}_s}(\mu _{s,mn,t}^{{\rm{line}}}) \le \rho _{s,mn,t}^{{\rm{line}}} \\
            \text{E}_{\mathbb{P}_s}(\mu _{s,k}^{{\rm{load}}}) \le \rho_{s,k}^{{\rm{load}}} 
        \end{array} \right.
    \end{array} 
    \right\}
\end{equation}
where $\mathbb{P}_s$ denotes the set of all probability distributions $\mathcal{P}(\Xi_s)$ on a $\sigma$-algebra of the support set $\Xi_s$. $\mu _{s,mn,t}^{{\rm{line}}}$ is the binary variable indicating whether branch $mn$ is on outage (1) or not (0); and $\mu _{s,k}^{{\rm{load}}}$ is the binary variable indicating whether user at bus $k$ would pay for the insurance (1) or not (0).

The first line in \eqref{eq: Ambiguity Set} indicates that all realizations of $\mu _{s,mn,t}^{{\rm{line}}}$ and $\mu _{s,k}^{{\rm{load}}}$ are within $\Xi_s$. While the second and third line mean that the first moment of line outages in extreme scenario $s$ and user buying insurance before extreme scenario $s$ subjected to their upper limits $\rho _{s,mn,t}^{{\rm{line}}}$ and $\rho _{s,k}^{{\rm{load}}}$, respectively. 

In order to facilitate the solution of the SDRO, the maximum number of line outages in the same period are limited to $N_{\rm{out}}^{\rm{line}}$ in this work, i.e., $\Xi_s$ is denoted according to $N-k$ contingencies. Besides, considering the limitation of DN resilience, only a part of users, with upper limitation to $N^{\rm{load}}_{\rm{ins}}$ in this work, can be provided with resilience insurance. Accordingly, the support set $\Xi_s$ can be defined as follow:
\begin{equation}
    \Xi_s = \left\{  
    \begin{array}{ll}
        (\mu _{s,mn,t}^{{\rm{line}}}, \ \mu _{s,k}^{{\rm{load}}}) \left\lvert\, \begin{array}{l}
            \sum_{mn} \mu _{s,mn,t}^{{\rm{line}}} \le N_{\rm{out}}^{\rm{line}}  \\
            \sum_{k} \mu _{s,k}^{{\rm{load}}} \le N_{\rm{ins}}^{\rm{load}}
        \end{array} \right.
    \end{array} 
    \right\}
\end{equation}

In addition, the line outage probability $\rho _{s,mn,t}^{{\rm{line}}}$ in \eqref{eq: Ambiguity Set} can be statistically obtained from historical data. While the The probability of users purchasing resilience insurance before extreme events $\rho _{s,k}^{{\rm{load}}}$ can be determined by \eqref{eq: rho load}.

\subsection{Objective Function} 
The objective function $F_{\rm{obj}}$ is to maximize the operator's net income $C_{\rm{NET}}$, as show in \eqref{eq: obj}, which consists of two parts: the cost of E-SOP investment $C_{\rm{INV}}$ and the income from resilience insurance services $C_{\rm{INS}}$.

\begin{figure*}[!t]
\normalsize
\setcounter{MYtempeqncnt}{\value{equation}}
\setcounter{equation}{\value{MYtempeqncnt}}
\begin{equation}    \label{eq: obj}
    F_{\rm{obj}} = \underbrace{ \max\limits_{\bm{x}, \bm{\alpha}} \ -C_{\rm{INV}} + \underbrace{ \sum\limits_s p_s \underbrace{ \min\limits_{\mathbb{P}_s \in \mathcal{B}} \underbrace{ \text{E}_{\mathbb{P}_s}  \left[ C_{{\rm{INS}},s}(\bm{x},\bm{\mu}_s) \right] }_{\text{layer IV: determine optimal operation scheme.}} }_{\text{layer III: determine "worst" line outage and insurance purchase statue.} } }_{\text{layer II: consider different extreme events of varying intensity.}} }_{\text{layer I: determine optimal allocation scheme and the optimal insurance pricing strategy.}}
\end{equation}
\stepcounter{MYtempeqncnt}
\setcounter{equation}{\value{MYtempeqncnt}}
\hrulefill
\vspace*{4pt}
\end{figure*}

In \eqref{eq: obj}, $\bm{x}$, $\bm{\alpha}$ and $\bm{\mu_s}$ are the vectors formed by all the variables related to the E-SOP planning, insurance compensation and premium for all scenarios and uncertainty variables under scenario $s$; $p_s$ donates the probability of extreme event scenarios of different intensities; $\text{E}_{\mathbb{P}_s} (\bullet)$ represents the expectation of $\bullet$ under PDF $\mathbb{P}_s$. $C_{\rm{INV}}$ and $C_{{\rm{INS}},s}$ in \eqref{eq: obj} are shown as follows:
\begin{align}
    \begin{split}
        & C_{\rm{INV}} = \frac{{r{{(1 + r)}^{y_r}}}}{{{{(1 + r)}^{y_r}} - 1}} \left( \sum\nolimits_{j} {c^{{\rm{SOP}}}}S_j^{{\rm{SOP}}} \right. \\
        & \left. \qquad \qquad \qquad \qquad \qquad + c_{\rm{e}}^{{\rm{ESS}}}{E^{{\rm{ESS}}}} + c_{\rm{p}}^{{\rm{ESS}}}{P^{{\rm{ESS}}}} \right)
    \end{split} \\
    \begin{split}
        & C_{{\rm{INS}},s} = \max\limits_{\bm{y}_s} \left( \alpha^{\rm{PRE}}_s \sum\nolimits_j \sum\nolimits_t \mu_{s,j}^{\rm{load}} P_{s,j,t}^{\rm{load}} \right.    \\
        & \left. \qquad \qquad \qquad \quad - \alpha^{\rm{COM}}_s \sum\nolimits_j \sum\nolimits_t P_{s,j,t}^{\rm{load,loss}} \right)
    \end{split}
\end{align}
where $r$ is the discount rate; $y_r$ is the lifespan of E-SOP; $c^{{\rm{SOP}}}$, $c_{\rm{e}}^{{\rm{ESS}}}$ and $c_{\rm{p}}^{{\rm{ESS}}}$ stand for investment cost per unit of SOP, per unit of ESS power and per unit of ESS capacity, respectively; $P^{\rm{ESS}}$ and $E^{\rm{ESS}}$ are ESS's installation power and capacity; $S_j^{{\rm{SOP}}}$ is the installation capacity of SOP converter in bus $j$; $\bm{y}_s$ is the vector formed by all the variables related to the DN operation under scenario $s$; $P_{s,j,t}^{{\rm{load}}}$ and $P_{s,j,t}^{{\rm{load,loss}}}$ are the demand and shut power of load.

In \eqref{eq: obj}, we develop a four-layer SDRO objective function and the meaning of each layer is indicated by the subscript. In this SDRO model, the uncertainty in the intensity of contingent extreme events is addressed by a stochastic optimization approach (see layer II), while the uncertainty in the occurrence of outages and resilience insurance purchases resulting from a specific extreme event is addressed via a distributionally robust optimization approach (see layer III). 

\subsection{Constrains}
\subsubsection{Insurance Constrains}
Insurance premiums and compensation need to be constrained within a reasonable range:
\begin{equation}
    0 \le \alpha^{\rm{PRE}}_s \le \alpha^{\rm{COM}}_s \le \alpha_{\max}
\end{equation}

\subsubsection{Investment Constrains}
The allocation scheme needs to meet the following constraints:
\begin{align}
    & {L^{{\rm{ESS}}}}P_{{\rm{min}}}^{{\rm{ESS}}} \le {P^{{\rm{ESS}}}} \le {L^{{\rm{ESS}}}}P_{\max }^{{\rm{ESS}}} \label{eq: ESS P max con} \\ 
    & {L^{{\rm{ESS}}}}E_{{\rm{min}}}^{{\rm{ESS}}} \le {E^{{\rm{ESS}}}} \le {L^{{\rm{ESS}}}}E_{\max }^{{\rm{ESS}}}, \ &  L^{{\rm{ESS}}} \in \{ 0,1\} \\
    & L_j^{{\rm{SOP}}}S_{{\rm{min}}}^{{\rm{SOP}}} \le S_j^{{\rm{SOP}}} \le L_j^{{\rm{SOP}}}S_{\max }^{{\rm{SOP}}} \label{eq: SOP S max con} \\
    & \sum\nolimits_{j} L_j^{{\rm{SOP}}} = N^{{\rm{SOP}}}, \  & L_j^{{\rm{SOP}}} \in \{ 0,1\}  \label{eq: SOP terminal con} \\
    & 0 \le {C_{{\rm{inv}}}} \le {C_{{\rm{inv,max}}}} \label{eq: max inv con}
\end{align}
where $L^{{\rm{ESS}}}$ and $L_j^{{\rm{SOP}}}$ are the binary variables indicating whether ESS/SOP is equipped (1) or not (0); $N^{{\rm{SOP}}}$ is the number of SOP ports; $C_{{\rm{inv,max}}}$ is the upper limit of budget.

\subsubsection{Operation Constrains}
The DN with E-SOP need to meet the following constraints during operation, including ESS constraints, SOP constraints, and other constraints of the DN. 

The ESS's model are shown as follow.
\begin{align}
    & 0 \le P_{{\rm{ch}},s,t}^{{\rm{ESS}}} \le P^{{\rm{ESS}}}, \quad  0 \le P_{{\rm{dis}},s,t}^{{\rm{ESS}}} \le P^{{\rm{ESS}}}  \label{eq: ESS charge con} \\
    & \xi _{\min }^{{\rm{SOC}}}E^{{\rm{ESS}}} \le \xi _{s,t}^{{\rm{SOE}}} \le \xi _{\max }^{{\rm{SOC}}}E^{{\rm{ESS}}} \label{eq: ESS SOE con} \\
    & \xi _{s,t}^{{\rm{SOE}}} = \xi _{s,t - 1}^{{\rm{SOE}}} + (P_{{\rm{ch}},s,t}^{{\rm{ESS}}}\eta^{{\rm{ESS}}} - P_{{\rm{dis}},s,t}^{{\rm{ESS}}}/\eta^{{\rm{ESS}}})\Delta T \label{eq: ESS SOE transfer con} \\
    & \xi _{s,1}^{{\rm{SOE}}} = \xi _{s,T}^{{\rm{SOE}}} \label{eq: ESS SOE return con}
\end{align}
where $P_{{\rm{ch}},s,t}^{\rm{ESS}}$ and $P_{{\rm{dis}},s,t}^{\rm{ESS}}$ are the charge and discharge power of the ESS; $\xi _{s,t}^{{\rm{SOE}}}$ stands for the SOE and $\xi _{\min }^{{\rm{SOC}}}$ and $\xi _{\max }^{{\rm{SOC}}}$ represent the minimum and maximum value of state of charge (SOC); $\eta^{\rm{ESS}}$ denotes ESS's charge/discharge efficiency.

While SOP’s model contain the following constrains.
\begin{align}
    & \sum\nolimits_{j} P_{s,j,t}^{{\rm{SOP}}} = P_{{\rm{dis}},s,t}^{{\rm{ESS}}} - P_{{\rm{ch}},s,t}^{{\rm{ESS}}} \label{eq: SOP zero net power con} \\
    &  - {\mu ^{{\rm{SOP}}}}S_j^{{\rm{SOP}}} \le Q_{s,j,t}^{{\rm{SOP}}} \le {\mu ^{{\rm{SOP}}}}S_j^{{\rm{SOP}}} \label{eq: SOP converter reactive power con} \\
    & {\left\| {{{[P_{s,j,t}^{{\rm{SOP}}} \quad Q_{s,j,t}^{{\rm{SOP}}}]}^{\rm{T}}}} \right\|_2} \le S_{j}^{{\rm{SOP}}}  \label{eq: SOP converter capacity con}
\end{align}
where $P_{s,j,t}^{{\rm{SOP}}}$ and $Q_{s,j,t}^{{\rm{SOP}}}$ stand for the active and reactive power of SOP injected into bus $j$, respectively; $\mu ^{{\rm{SOP}}}$ is the reactive power limitation coefficient; $\left\| \bullet \right\|_2$ denotes the 2-norm of $\bullet$.

In addition, the DN also needs to meet the following constraints, including second-order cone power flow constraints, branch outage constraints and safety constraints, see (9)-(10) and (12) in \cite{liji} for detail.

\subsection{Solution Method}
The proposed SDRO model can be transformed into the following matrix form:
\begin{equation}
    \begin{split}
        & \min\limits_{\bm{x}, \bm{\alpha}} \ {\bm{c}}^{\rm{T}}\bm{x} + \sum\limits_s p_s \sup\limits_{\mathbb{P}_s \in \mathcal{B}} \text{E}_{\mathbb{P}_s} \left[ Q(\bm{x},\bm{\mu}_s) \right] \\
        & {\rm{s.t.}} \ \bm{Ax} \le \bm{b}, \qquad \bm{D\alpha} \le \bm{e}
    \end{split}
\end{equation}
\begin{equation}
    \mathcal{B} = \left\{  
    \begin{array}{ll}
        \mathbb{P}_s \in \mathcal{P}(\Xi_s) \left\lvert\, \begin{array}{l}
            \Pr(\bm{\mu}_s \in \Xi_s) = 1, \\
            \text{E}_{\mathbb{P}_s}(\bm{\mu}_s) \le \bm{\rho}_{s} + \bm{J} \bm{\alpha}
        \end{array} \right.
    \end{array} 
    \right\}
\end{equation}
\begin{equation}
    \Xi_s = \left\{  
    \begin{array}{ll}
        \bm{\mu}_s \left\lvert\, 
            \bm{K} \bm{\mu}_s \le \bm{g} \right.
    \end{array} 
    \right\}
\end{equation}
\begin{equation}    \label{eq: Q}
    \begin{split}
        Q(\bm{x}, \bm{\mu}_s) = & \min\limits_{\bm{y}_s} \ {\bm{d}}^{\rm{T}}\bm{y}_s \\
        & {\rm{s.t.}} \ \bm{E}\bm{y}_s  \le \bm{f} - \bm{G}\bm{\mu}_s - \bm{H}\bm{x}, \\
        & \quad \left\| {\bm{B}{\bm{y}}_s} \right\|_2 \le {\bm{r}^{\rm{T}}}{\bm{y}_s}
    \end{split}
\end{equation}


Applying the definition of expectation and strong duality theory, the original SDRO can be converted into the following form, with the detailed derivation process described in \cite{zhouwei}.
\begin{equation}
    \begin{split}
        & \min\limits_{\bm{x}, \bm{\alpha}, \bm{\zeta}_s, \bm{\beta}_s \ge 0} \ {\bm{c}}^{\rm{T}}\bm{x} + \sum\limits_{s} p_s \left[ \bm{\zeta}_s + \bm{\beta}_s^{\rm{T}} (\bm{\rho}_{s} + \bm{J} \bm{\alpha}) \right]    \\
        & {\rm{s.t.}} \ \bm{Ax} \le \bm{b}, \qquad \bm{D\alpha} \le \bm{e}  \\
        & \qquad \bm{\zeta}_s \ge \max_{\bm{\mu}_s \in \mathcal{B}} \left[ Q(\bm{x},\bm{\mu}_s) - \bm{\beta}_s^{\rm{T}} \bm{\mu}_s \right]
    \end{split}
\end{equation}
where $\bm{\zeta}_s$ and $\bm{\beta}_s$ are auxiliary variables, and the definition of $Q(\bm{x},\bm{\mu}_s)$ is consistent with \eqref{eq: Q}.

The SDRO model is transformed into a min-max-min problem, which can be solved by C\&CG algorithm. The specific steps of CC\&G algorithm are shown in Section 3.3 of \cite{zhouli}.

\section{Case Study}
\label{sec: Case Study}
\begin{figure}[ht]
\centerline{\includegraphics[width=0.9\columnwidth]{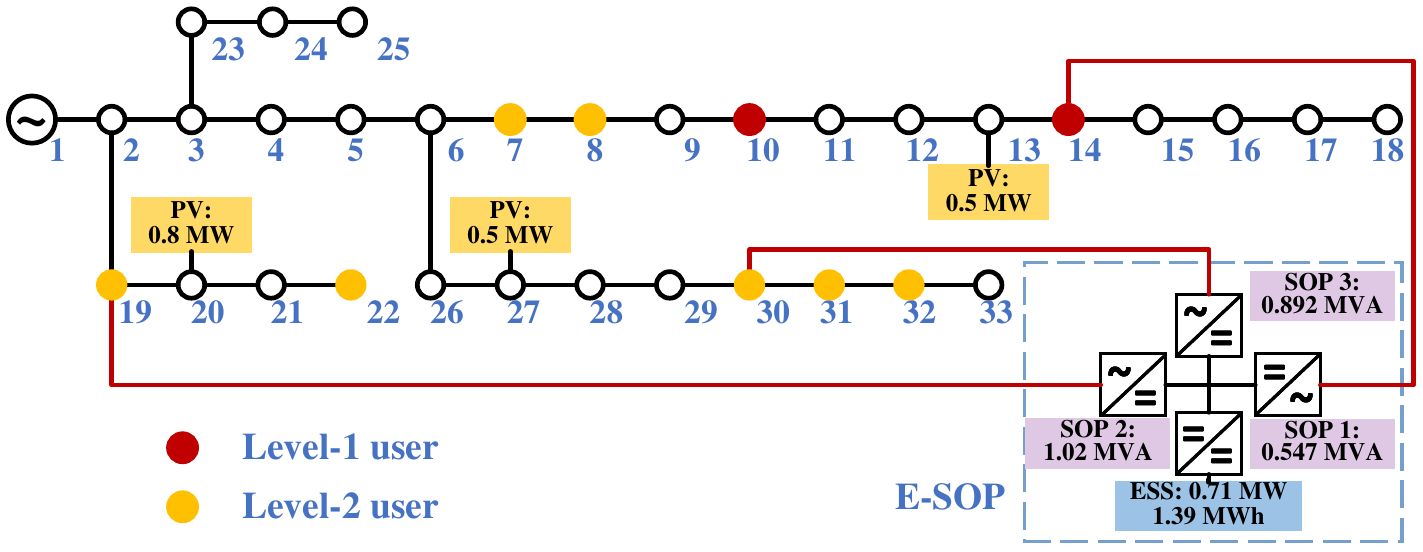}}
\caption{Modified IEEE 33-bus system with E-SOP.}
\label{fig: Modified IEEE 33-bus system with E-SOP}
\end{figure}

The proposed model is applied to a modified IEEE 33-bus system as shown in Fig. \ref{fig: Modified IEEE 33-bus system with E-SOP}. Bus 10 and 14 are treated as level-1 loads, while bus 7, 8, 24, 25, 30, 31 and 32 are considered as level-2 users, and others are considered as level-3 loads. bus 13, 20 and 27 are connected to 0.5 MW, 0.8 MW and 0.5 MW PV, respectively. The historical data of PV's output and load demand is obtained from \cite{zhouzhang}.

In terms of scenario generation, we select typhoon as a typical extreme event and assume that there are 3 typhoons per year, each lasting 3 days. 6 typhoon scenarios are set corresponding to different typhoon classes, namely, tropical depression, tropical storm, severe tropical storm, typhoon, severe typhoon and super typhoon, with the occurrence probability of 16.1\%, 24.4\%, 21.7\%, 19.1\%, 10.7\% and 8.0\% respectively \cite{yang}.

The optimal configuration scheme of E-SOP is also shown in Fig.\ref{fig: Modified IEEE 33-bus system with E-SOP}. SOPs are connected to bus 14, 19 and 30, with capacities of 0.547 MVA, 1.02 MVA and 0.892 MVA, respectively. The configured ESS is 0.71 MW / 1.39 MWh. The total investment cost of the E-SOP is 4 606 350 CNY. In each scenario, the insurance premium, compensation and system operator's profit are shown in Table \ref{table: Premium, compensation and profit of each scenario}.

\begin{table}[!h]
    \caption{Premium, compensation and profit of each scenario}
    \label{table: Premium, compensation and profit of each scenario}
    \centering
    \footnotesize
    \begin{tabular}{cccc}
        \toprule
        Typhoon classes & \makecell{Premium\\(CNY / kWh)} & \makecell{Compensation\\(CNY / kWh)} & \makecell{Profit\\(CNY)}    \\
        \midrule
        Tropical depression & 1.91 & 8.21 & 25 394        \\
        Tropical storm & 3.09 & 11.67 & 47 188      \\
        Severe tropical storm & 4.83 & 15.21 & 98 926      \\
        Typhoon & 6.60 & 19.22 & 130 016      \\
        Severe typhoon & 8.81 & 24.04 & 148 029      \\
        Super typhoon & 10.82 & 27.79 & 131 435      \\
        \bottomrule   
        \normalsize
    \end{tabular}
\end{table}

The E-SOP is assumed to have a usable life of 15 years. 
The investment cost will be recovered in the 12th year, with an internal rate of return (IRR) of 9.28\%. In other words, the resilience insurance can help E-SOP investors recover project costs and realize profits. Besides, E-SOP can also improve the flexibility of the DN under normal events, such as facilitating renewable energy integration and enabling bidirectional power flow control. Thus, the actual benefits of investing in the construction of E-SOP will be higher than the above analysis.

\begin{figure}
	\centering
	\subfigure[Load supply of all loads in the DN.]{
		\begin{minipage}{0.81\columnwidth}
                        \includegraphics[width=\textwidth]{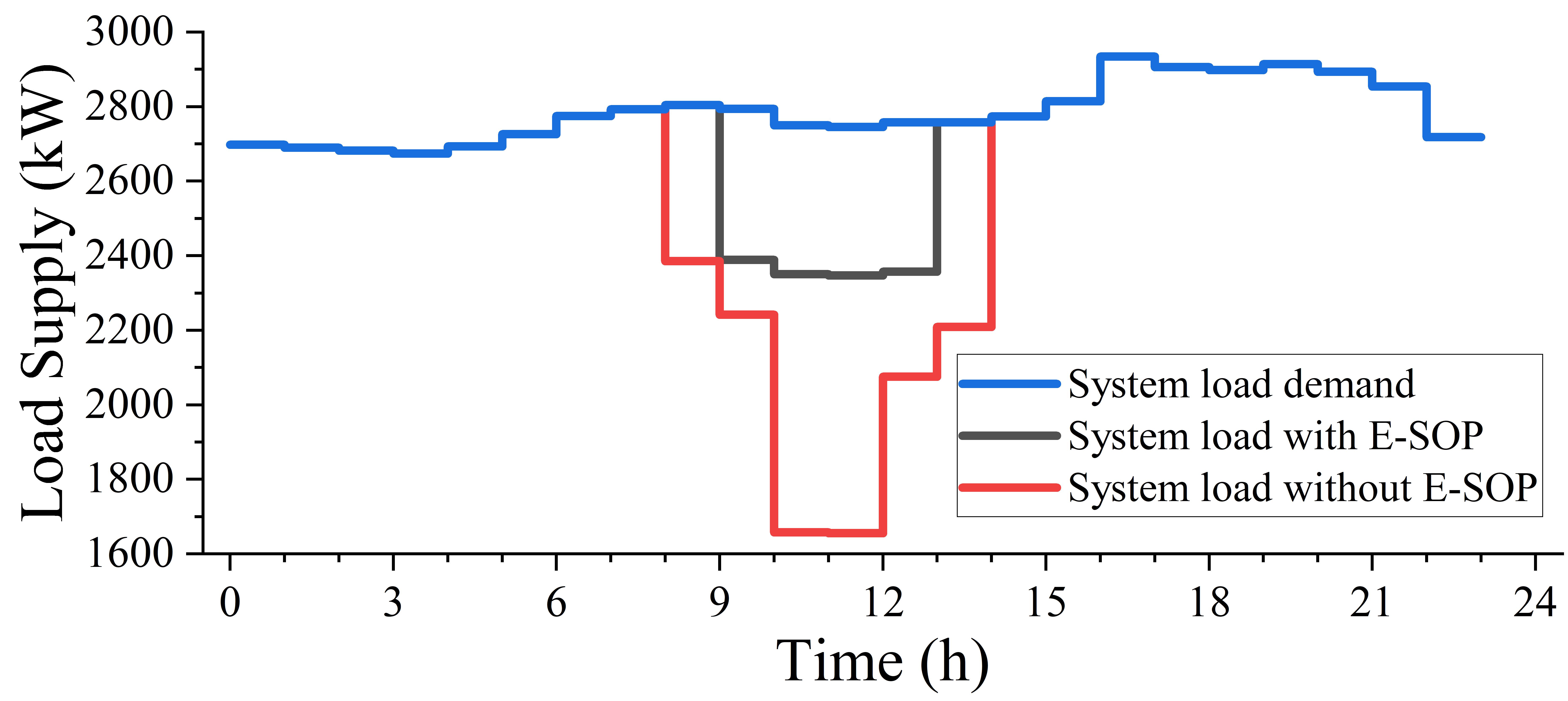} \\
		\end{minipage}    \label{subfig: Load supply of all loads in the DN}
	}

	\subfigure[Load supply of level-1 and level-2 users.]{
		\begin{minipage}{0.81\columnwidth}
			\includegraphics[width=\textwidth]{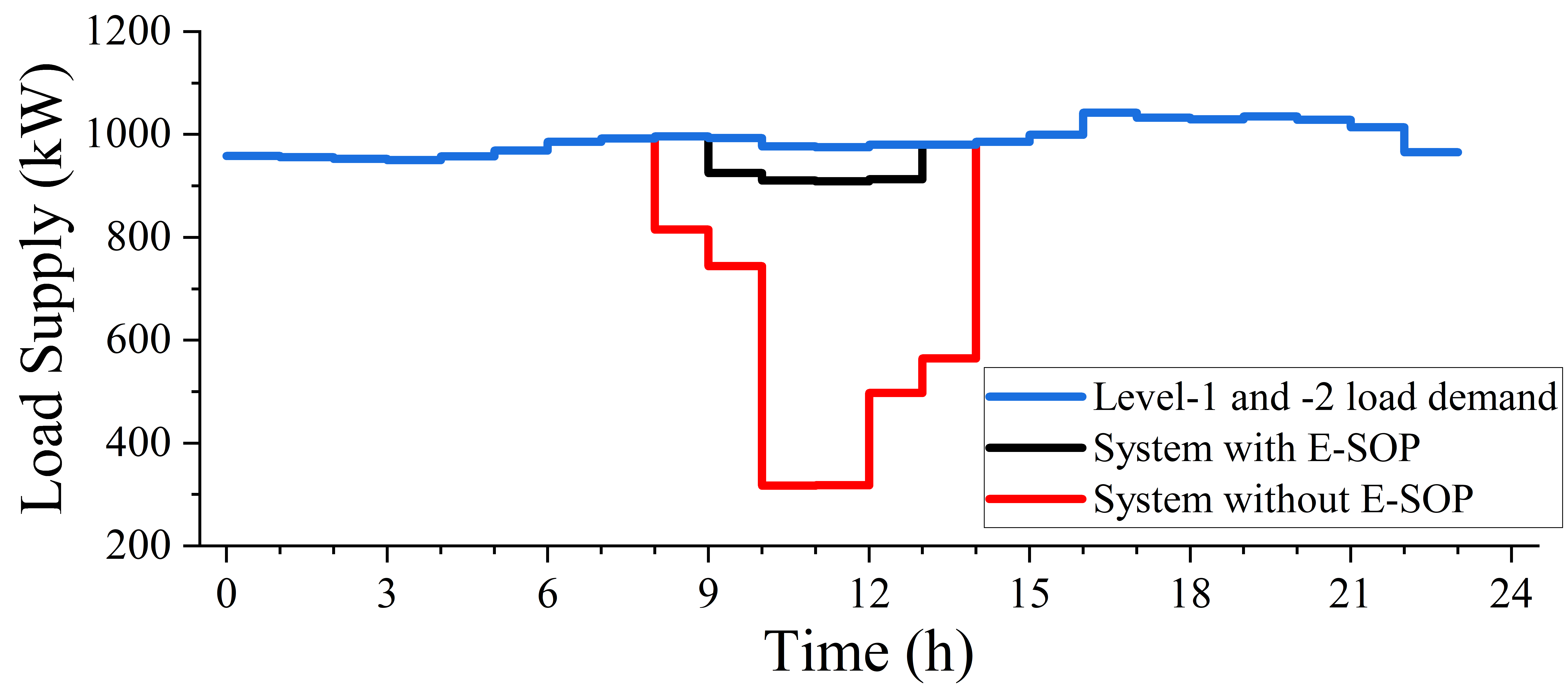} \\
		\end{minipage}    \label{subfig: Load supply of level-1 and level-2 users}
	}
 	\caption{Comparison of load supply with and without E-SOP.} 
	\label{fig: Comparison of load supply in typhoon scenario with and without E-SOP}
\end{figure}

We take the "typhoon" scenario as an example to analyze the effect of E-SOP on improving resilience. The load supply at each time is shown in Fig.\ref{subfig: Load supply of all loads in the DN}. Compared with the original DN, the load loss of the DN equipped with E-SOP is reduced by 63.44\%, and the minimum load power is increased by 29.46\%, indicating that the DN resilience is significantly improved. While as shown in Fig.\ref{subfig: Load supply of level-1 and level-2 users}, the introduction of E-SOP effectively protected Level-1 and Level-2 users. Compared to the benchmark system, the load shut of these key users decreased by 89.88\% in the system with E-SOP.

\section{Conclusion}
\label{sec: Conclusion}
This paper proposes an E-SOP cost allocation framework based on resilience insurance services and establishes an SDRO model to determine the optimal planning scheme of E-SOP and insurance pricing strategy. The numerical results show that the proposed cost allocation mechanism can help the DN operator recover the investment cost during the lifespan of E-SOP (IRR is 9.28\%). Besides, due to the introduction of E-SOP, the total load loss in the typhoon scenario is reduced by 63.44\%, and the loss of load for key users is reduced by 89.88\%, indicating that the resilience is significantly improved.

\section*{Acknowledgment}
This work was supported by the National Natural Science Foundation of China (U22B20103, 52307137).

\end{document}